\documentclass[showpacs,10pt,twocolumn,prb]{revtex4-1}
\usepackage{amsmath}
\usepackage{amssymb}
\usepackage{graphics}
\usepackage{epsfig}
\usepackage{CJK}
\usepackage{color}

\begin{document}


\title{Anomalous Hall effect of ferromagnetic Fe$_{3}$Sn$_{2}$ single crystal with geometrically frustrated kagome lattice}
\author{Qi Wang$^{1}$, Shanshan Sun$^{1}$, Xiao Zhang$^{2}$, Fei Pang$^{1}$, and Hechang Lei$^{1,}$}
\email{hlei@ruc.edu.cn}
\affiliation{$^{1}$Department of Physics and Beijing Key Laboratory of Opto-electronic Functional Materials $\&$ Micro-nano Devices, Renmin University of China, Beijing 100872, China}
\affiliation{$^{2}$State Key Laboratory of Information Photonics and Optical Communications $\&$ School of Science, Beijing University of Posts and Telecommunications, Beijing 100876, China.}
\date{\today}

\begin{abstract}

The anomalous Hall effect is investigated for ferromagnetic Fe$_{3}$Sn$_{2}$ single crystal with geometrically frustrated kagome bilayer of Fe. The scaling behavior between anomalous Hall resistivity $\rho_{xy}^{A}$ and longitudinal resistivity $\rho_{xx}$ is quadratic and further analysis implies that the AHE in Fe$_{3}$Sn$_{2}$ single crystal should be dominated by the intrinsic Karplus-Luttinger mechanism rather than extrinsic skew-scattering or side-jump mechanisms. Moreover, there is a sudden jump of anomalous Hall conductivity $\sigma_{xy}^{A}$ appearing at about 100 K where the spin-reorientation transition from the $c$ axis to the $ab$ plane is completed. This change of $\sigma_{xy}^{A}$ might be related to the evolution of Fermi surface induced by the spin-reorientation transition.

\end{abstract}


\maketitle


\section{Introduction}

In magnetic materials, in addition to the ordinary Hall effect originating from the deflection of moving charge carriers by the Lorentz force in a magnetic field, there is an anomalous term which is proportional to the spontaneous magnetization $M$. Although this anomalous Hall effect (AHE) has been recognized for more than a century,\cite{Hall} it is still a long-standing topic in condensed matter physics which attracted great interest due to the fundamental physics and great potential applications. It is now widely accepted that there are three mechanisms responsible for the AHE.\cite{Nagaosa} Karplus and Luttinger first proposed a model related to the band structure of ferromagnetic metals with the spin-orbit interaction (SOI) to explain the AHE (intrinsic KL mechanism).\cite{Karplus} In modern language, it has been reinterpreted as a manifestation of Berry-phase effects on occupied electronic Bloch states.\cite{Jungwirth,Onoda1} For this mechanism, the anomalous Hall resistivity ($\rho_{xy}^{A}$) scales quadratically with longitudinal resistivity ($\rho_{xx}$). The AHE can also be induced by asymmetric scattering of conduction electrons which are subject to SOI. These extrinsic mechanisms for the AHE involve skew-scattering and side-jump scattering ones.\cite{Smit,Berger} For the former, the $\rho_{xy}^{A}$ is linearly proportional to $\rho_{xx}$. By contrast, the latter gives $\rho_{xy}^{A}\propto\rho_{xx}^{\alpha}$ ($\alpha=$ 2), similar to the intrinsic mechanism.

Recently, the AHE of materials with frustrated structure and/or non-collinear magnetism, such as spin-ice like ferromagnetic Nd$_{2}$Mo$_{2}$O$_{7}$,\cite{Taguchi} Pr$_{2}$Ir$_{2}$O$_{7}$ with chiral spin-liquid state,\cite{Machida} and non-collinear antiferromagnetic Mn$_{3}$Sn,\cite{Nakatsuji} have attracted much attention. These systems exhibit large AHE like in the traditional ferromagnet possibly due to the effects of Berry-phase originating from the non-coplanar spin sites with spin chirality. These studies deepen our understanding of the relation between topological nature of the quantum states of matter and its transport properties, however, the effects of frustrated magnetism on the AHE, i.e., the interaction between localized spins in the frustrated lattice and itinerant electrons, are still not fully solved because of limited materials systems and theoretical calculations until now.

Ferromagnetic Fe$_{3}$Sn$_{2}$ with geometrically frustrated kagome bilayer of Fe is a good candidate to study this issue. Previous neutron powder diffraction (NPD) and M\"{o}ssbauer spectra show that the Curie temperature of Fe$_{3}$Sn$_{2}$ polycrystal is about 640 K and it may be a frustrated ferromagnet with a temperature-dependent non-collinear spin structure.\cite{Malaman,Caer1,Caer2,Fenner} More surprisedly, Fe$_{3}$Sn$_{2}$ polycrystal exhibits AHE with unusual scaling relationship between the anomalous Hall coefficient $R_{s}$ and the longitudinal resistivity $\rho_{xx}$, $R_{s}\propto\rho_{xx}^{3.15}$, i.e., $\alpha=$ 3.15 when the $M$ is almost unchanged with temperature.\cite{Kida} This large scaling index can not be explained by either one of three mechanisms introduced above. Moreover, theoretical calculation suggests that Fe$_{3}$Sn$_{2}$ could be a model system to exhibit flat-band ferromagnetism, leading to the fractional quantum Hall states at high temperatures (maybe even room temperatures).\cite{Tang} These results motivate us to study the AHE of Fe$_{3}$Sn$_{2}$ in detail. In this work, we report the results of transport and magnetic properties of Fe$_{3}$Sn$_{2}$ single crystal. Distinctly different from $\alpha=$ 3.15 in polycrystal, the quadratically scaling behavior between anomalous Hall resistivity $\rho_{xy}^{A}$ and $\rho_{xx}$ are observed in Fe$_{3}$Sn$_{2}$ single crystal and our analysis implies that the main contribution of observed AHE should originate from the intrinsic KL mechanism. The large Hall conductivity might be intimately related to the topologically nontrivial bands in Fe$_{3}$Sn$_{2}$, originating from the ferromagnetism with geometrically frustrated kagome bilayer of Fe.

\section{Experimental}

Single crystals of Fe$_{3}$Sn$_{2}$ were grown by the self-flux method with Fe : Sn = 4 : 96 molar ratio. Fe (purity 99.98 \%) and Sn (purity 99.99 \%) grains were put into alumina crucible and sealed in quartz ampoule under partial argon atmosphere. The sealed quartz ampoule was heated to and soaked at 1323 K for 4 h, then cooled down to 1043 K with 6 K/h. At this temperature, the ampoule was taken out from the furnace and decanted with a centrifuge to separate Fe$_{3}$Sn$_{2}$ crystals from Sn flux. X-ray diffraction (XRD) of a single crystal was performed using a Bruker D8 X-ray machine with Cu $K_{\alpha}$ radiation ($\lambda=$ 0.15418 nm) at room temperature. Magnetization and electrical transport measurements were carried out by using Quantum Design PPMS-9 and MPMS-XL5. The longitudinal and Hall electrical resistivity were performed using a four-probe method on single crystals cutting into rectangular shape with the dimensions of 1.9$\times$1.5$\times$0.4 mm$^{3}$. The current flows in the $ab$ plane of hexagonal structure. For the measurement of Hall resistivity $\rho_{xy}$, in order to effectively get rid of the longitudinal resistivity contribution due to voltage probe misalignment, the Hall resistivity was obtained by the difference of transverse resistance measured at the positive and negative fields, i.e., $\rho_{xy}(\mu_{0}H)=[\rho(+\mu_{0}H)-\rho(-\mu_{0}H)]/2$.

\section{Results and discussion}

\begin{figure}[tbp]
\centerline{\includegraphics[scale=0.4]{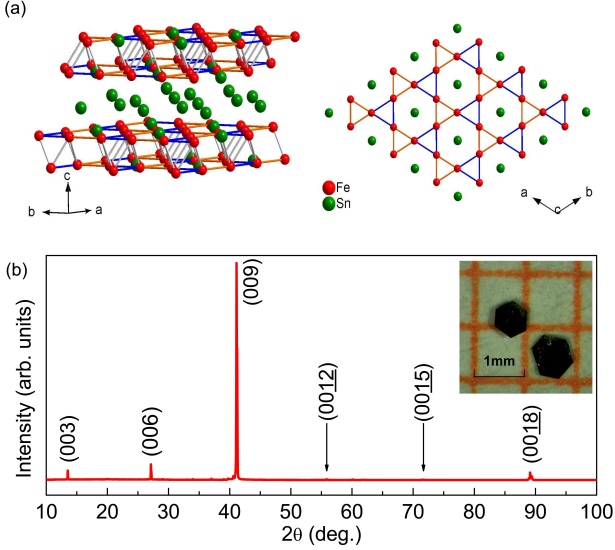}} \vspace*{-0.3cm}
\caption{(a) Crystal structure of Fe$_{3}$Sn$_{2}$ (left) and kagome layer of Fe atoms (right). The small red and big green balls represent Fe and Sn atoms, respectively. (b) XRD pattern of a Fe$_{3}$Sn$_{2}$ single crystal. Inset: photo of typical Fe$_{3}$Sn$_{2}$ single crystals.}
\end{figure}

As shown in Fig. 1(a), the crystal structure of Fe$_{3}$Sn$_{2}$ is composed of Fe-Sn bilayer and Sn layer stacking along the $c$-axis direction alternatively. In Fe-Sn bilayer, the Fe atoms form two kagome layers with two kinds of equilateral triangles which have different Fe-Fe bond lengths (0.2732 and 0.2582 nm for orange and blue ones, respectively). The Sn atoms occupy the centers of the hexagons in the kagome layers. Moreover, the triangles of Fe atoms with small side length in two layers connect each other and make up the octahedra of Fe atoms. On the other hand, there is another Sn layers in between two Fe-Sn bilayer. The Sn atoms in this layer form a two-dimensional network of six-member ring connected each other by edge-sharing, which is same as in graphene. Fig. 1(b) shows the XRD pattern of a Fe$_{3}$Sn$_{2}$ single crystal. All of peaks can be indexed by the indices of (0 0 l) lattice planes. It indicates that the crystal surface is parallel to the $ab$ plane and perpendicular to the $c$-axis. The inset of Fig. 1(b) shows a photograph of typical Fe$_{3}$Sn$_{2}$ crystals on a 1 mm grid paper. It can be seen that these crystals have hexagonal shape with small thickness, consistent with the layered structure and the rhombohedral symmetry of Fe$_{3}$Sn$_{2}$ with space group $R\bar{3}m$.

\begin{figure}[tbp]
\centerline{\includegraphics[scale=0.42]{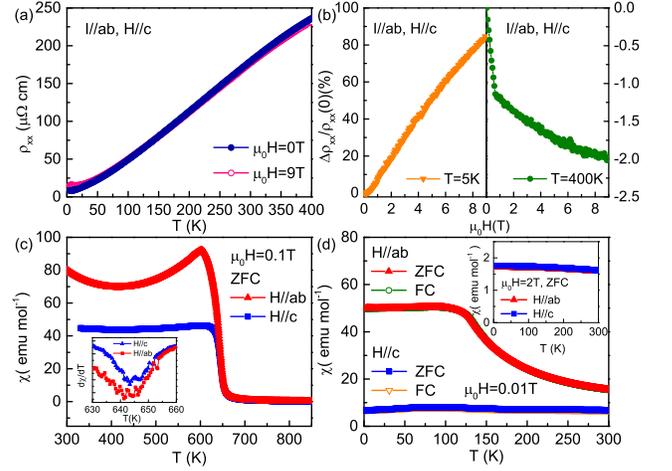}} \vspace*{-0.3cm}
\caption{(a) Temperature dependence of the in-plane resistivity $\rho_{xx}(T)$ of Fe$_{3}$Sn$_{2}$ single crystal at zero field and 9 T along the $c$ axis. (b) Magnetoresistance of $\rho_{xx}(\mu_{0}H)$ at 5 and 400 K for $H\Vert c$. (c) Temperature dependence of magnetic susceptibility $\chi(T)$ for Fe$_{3}$Sn$_{2}$ single crystal in the high-temperature region (300 - 850 K) with ZFC mode at $\mu_{0}H=$ 0.1 T for $H\Vert ab$ and $H\Vert c$. Inset: corresponding $d\chi(T)/dT$ vs. $T$. (d) Temperature dependence of $\chi(T)$ with ZFC and FC modes at $\mu_{0}H=$ 0.01 T for both field directions. Inset: $\chi(T)$ vs. $T$ with ZFC mode at $\mu_{0}H=$ 2 T for both field directions.}
\end{figure}

Electrical resistivity $\rho_{xx}(T)$ in the $ab$ plane as a function of temperature for Fe$_{3}$Sn$_{2}$ single crystal is shown in Fig. 2(a). The $\rho_{xx}(T)$ at zero field exhibits metallic behavior with much smaller absolute value than that of polycrystal,\cite{Kida} partially due to the absence of grain boundary (GB) effects. The residual resistivity ratio (RRR) that $\rho_{xx}(300K)/\rho_{xx}(5K)$ is about 21, indicating the high quality of sample. At $\mu_{0}H=$ 9 T, the metallic behavior of $\rho_{xx}(T)$ is almost unchanged (Fig. 2(a)). But the signs of magnetoresistance (MR $=(\rho_{xx}(\mu_{0}H)-\rho_{xx}(0))/\rho_{xx}(0)=\Delta\rho_{xx}/\rho_{xx}(0)$) at low and high temperatures are opposite (Fig. 2(b)).The MR at 5 K is positive and reaches about 80 \% at $\mu_{0}H=$ 9 T. In contrast, there is a negative MR with small value ($\sim$ 2 \% at $\mu_{0}H=$ 9 T) at 400 K. The positive MR at low temperature should be originate from the Lorenz force caused by magnetic field on the carrier motion. On the other hand, The negative MR becomes obvious at high temperature near $T_{c}$, because of the suppression of spin scattering in ferromagnetic Fe$_{3}$Sn$_{2}$ by applying a magnetic field. Moreover, it can be seen that there is a slope change of $\rho_{xx}(\mu_{0}H)$ at $\mu_{0}H\sim$ 0.7 T, because the magnetization of Fe$_{3}$Sn$_{2}$ becomes saturated when field further increases. In other words, once the magnetic moment is saturated, further increase of field will have much weaker influence on the alignment of spin, the change of MR with field would also become smaller.

High-temperature magnetic susceptibility $\chi_{ab}(T)$ and $\chi_{c}(T)$ with zero-field-cooling (ZFC) mode at $\mu_{0}H=$ 0.1 T for $H\Vert ab$ and $H\Vert c$ are shown in Fig. 2(c). There are sharp increases for both $\chi_{ab}(T)$ and $\chi_{c}(T)$, corresponding to the ferromagnetic transition in Fe$_{3}$Sn$_{2}$. The Curies temperature $T_{c}$ determined from the peaks of $d\chi_{ab}(T)/dT$ and $d\chi_{c}(T)/dT$ is about 641 and 643 K, respectively (inset of Fig. 2(c)). It is very close to the $T_{c}$ of Fe$_{3}$Sn$_{2}$ polycrystal.\cite{Fenner} Surprisedly, when temperature is below $T_{c}$, the $\chi(T)$ start to decrease with temperature for both field directions at beginning and then increase when $T<T^{*}$ ($\sim$ 423 K). This behavior is also observed in Fe$_{3}$Sn$_{2}$ polycrystal with slightly different $T^{*}$ and it was ascribed to the rotation of the Fe moments from the $c$ axis to the $ab$ plane with decreasing temperature, consistent with previous results of NPD and M\"{o}ssbauer spectra.\cite{Malaman,Caer1,Caer2,Fenner} When $T<$ 100 K, the ZFC and field-cooling (FC) $\chi(T)$ curves for both field directions at $\mu_{0}H=$ 0.01 T (Fig. 2(d)) still almost overlap with only slight differences between each other. It is different from the obviously branching behavior between ZFC and FC $\chi(T)$ curves of Fe$_{3}$Sn$_{2}$ polycrystal in which there is a spin glass (SG) state appearing below 80 K.\cite{Fenner} The much weaker branching behavior in single crystal than in polycrystal implies that there may be a frustrated state rather than SG state because of the spin disorder at GBs in polycrystal might have significant influence on magnetic properties at low temperature. On the other hand, the $\chi_{c}(T)$ exhibits more obviously downturn behavior, which could be ascribed to the accelerated process of spin reorientation from the $c$ axis to the $ab$ plane.\cite{Caer2} When field increases to 2 T, both $\chi_{ab}(T)$ and $\chi_{c}(T)$ exhibit typical ferromagnetic behavior in the whole temperature range (inset of Fig. 2(d)), similar to previous results.\cite{Caer1} It implies that the spin reorientation process is suppressed by the high magnetic field.

\begin{figure}[tbp]
\centerline{\includegraphics[scale=0.35]{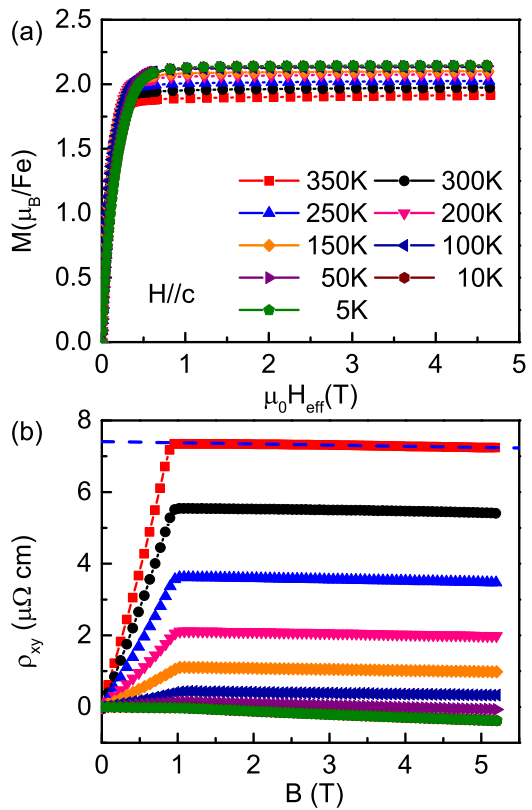}} \vspace*{-0.3cm}
\caption{(a) Effective field dependence of magnetization $M(\mu_{0}H_{\rm eff})$ and (b) Hall resistivity $\rho_{xy}(B)$ as a function of magnetic induction $B$ for Fe$_{3}$Sn$_{2}$ single crystal at various temperatures when $H\Vert c$. The blue dashed line in (b) is the linear fit of $\rho_{xy}(B)$ at high-$B$ region when $T=$ 350 K.}
\end{figure}

Fig. 3(a) exhibits the effective field dependence of magnetization $M(\mu_{0}H_{\rm eff})$ at various temperatures between 5 and 350 K for $H\Vert c$. Here $\mu_{0}H_{\rm eff}=\mu_{0}(H-N_{d}M)$, where $N_{d}$ is the demagnetizing factor and the calculated value of $N_{d}$ is 0.65. Because of $T<T_{c}$, the shape of $M(\mu_{0}H_{\rm eff})$ curves is typical for ferromagnets, i.e., a rapid increase at low field region with a saturation in higher magnetic fields because most of Fe moments have been aligned along the field direction. The saturation magnetization $M_{s}$ increases with decreasing temperature, well consistent with the temperature dependence of $\chi_{c}(T)$ curve at $\mu_{0}H=$ 2 T (inset of Fig. 2(d)) as well as the trend of magnetic moment obtained from NPD.\cite{Malaman} But it is in sharp contrast to the results of $M(\mu_{0}H_{\rm eff})$ for Fe$_{3}$Sn$_{2}$ polycrystal in which the $M_{s}$ becomes smaller with decreasing temperature. On the other hand, the $M_{s}$ at the lowest measuring temperature (5 K) for $H\Vert c$ is about 2.1 $\mu_{B}$/Fe, in good agreement with those values in the literature.\cite{Kida,Fenner,Malaman,Caer1}.

Hall resistivity $\rho_{xy}(B)$ as a function of magnetic induction $B$ for Fe$_{3}$Sn$_{2}$ single crystal at various temperatures is shown in Fig. 3(b). Here $B=\mu_{0}(H_{\rm eff}+M)=\mu_{0}(H+(1-N_{d})M)$. The $\rho_{xy}(B)$ increases quickly to certain saturated values at low $B$ region. With increasing $B$ further, the $\rho_{xy}(B)$ decreases slightly and the $B$ dependence of $\rho_{xy}(B)$ is almost linear, i.e., the $\rho_{xy}(B)/B$ is constant. On the other hand, when lowering temperature, the saturated value of $\rho_{xy}(B)$ becomes smaller gradually. These behaviors are qualitatively consistent with the results in Fe$_{3}$Sn$_{2}$ polycrystal, however, the saturated values at each temperature are much larger than those for polycrystal.\cite{Kida} This difference can be partially ascribed to the effects of GB scattering on $\rho_{xy}(B)$. Moreover, at 5 K, the inflecting behavior of $\rho_{xy}(B)$ becomes very small and the high-$B$ value is negative, suggesting that the dominant carrier in Fe$_{3}$Sn$_{2}$ is electron. It should be noted that the saturated value of $\rho_{xy}(B)$ for Fe$_{3}$Sn$_{2}$ single crystal at 350 K is enormous, much larger than that of typical itinerant ferromagnets, such as Fe and Ni.\cite{Volkenshtein,Kaul} There is a similarity between the shapes of $\rho_{xy}(B)$ and $M(\mu_{0}H_{\rm eff})$ curves. It clearly indicates that there is an AHE in itinerant ferromagnet Fe$_{3}$Sn$_{2}$.

In general, the Hall resistivity $\rho_{xy}$ in the ferromagnets is made up of two parts,\cite{Kida}

\begin{equation}
\rho_{xy} = \rho_{xy}^{O} + \rho_{xy}^{A} = R_{0}B+R_{s}\mu_{0}M
\end{equation}

where $\rho_{xy}^{O}$ and $\rho_{xy}^{A}$ is the ordinary and anomalous Hall resistivity, respectively. $R_{0}$ is the ordinary Hall coefficient from which apparent carrier concentration and type can be determined ($R_{0}\sim -1/|e|n_{a}$), and $R_{s}$ is the anomalous Hall coefficient. As shown in Fig. 3(b), the values of $R_{0}$ and $\rho_{xy}^{A}$ in principle can be determined from the linear fit of $\rho_{xy}(B)$ curve at the saturation region. The slope and $y$-axis intercept is corresponding to the $R_{0}$ and $\rho_{xy}^{A}$, respectively. The $R_{s}$ can be obtained by using the formula $\rho_{xy}^{A}=R_{s}\mu_{0}M_{s}$ with the $M_{s}$ taken from the $M(\mu_{0}H_{\rm eff})$ curves at $\mu_{0}H_{\rm eff}=$ 2 T (Fig. 3(a)). The $R_{0}(T)$ is negative at various temperature, in contrast to the positive $R_{s}(T)$ (Fig. 4(a) and (b)). The former confirms that the electron-type carrier is dominant. The $R_{0}(T)$ shows weak temperature-dependence above 100 K. When $T<$ 100 K, the absolute value of $R_{0}$ becomes larger. Using the values of $R_{0}$, the $n_{a}$ is evaluated (inset of Fig. 4(a)). It is insensitive to the temperature ($\sim$ 1.8$\times$10$^{22}$ cm$^{-3}$, corresponding to 1.47 carriers per Fe$_{3}$Sn$_{2}$) when $T>$ 150 K and increases slightly with decreasing temperature down to 100 K. Then the $n_{a}(T)$ reduces sharply with further lowering temperature below 100 K. The strong variation of $n_{a}(T)$ below 100 K is possibly due to the influence of spin reorientation on Fermi surface, leading to the change of carrier concentration. On the other hand, the $R_{s}(T)$ decreases with decreasing temperature monotonically and approaches to almost zero at low temperature (Fig. 4(b)). The value of $R_{s}$ at 0.5$T/T_{c}$ ($\sim$ 320 K) is about 1$\times$10$^{-9}$ $\Omega$ cm G$^{-1}$, somewhat larger than Fe$_{3}$Sn$_{2}$ polycrystal.\cite{Kida} Importantly, at same reduced temperature, Fe$_{3}$Sn$_{2}$ single crystal has about two orders of magnitude large $R_{s}$ when compared to the conventional itinerant ferromagnets, such as pure Fe and Ni.\cite{Volkenshtein,Kaul}

\begin{figure}[tbp]
\centerline{\includegraphics[scale=0.4]{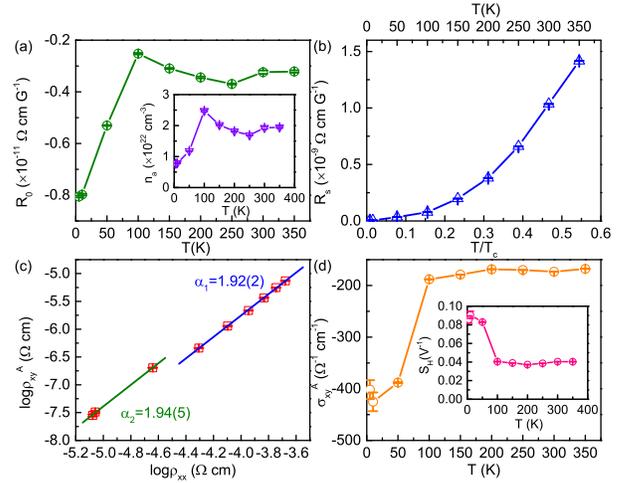}} \vspace*{-0.3cm}
\caption{(a) and (b) Temperature dependence of fitted $R_{0}(T)$ and $R_{s}(T)$ from $\rho_{xy}(B, T)$ curves using eq. (1). Inset of (a): derived $n_{a}(T)$ from $R_{0}(T)$. (c) Plot of log$\rho_{xy}^{A}(T)$ vs. log$\rho_{xx}(T)$. The blue and green solid lines are the fits using the relation $\rho_{xy}^{A}\propto \rho_{xx}(T)^{\alpha}$ below and above 100 K, respectively. (d) Anomalous Hall conductivity $\sigma_{xy}^{A}(T)$ as a function of temperature. Inset: temperature dependence of $S_{H}(T)$.}
\end{figure}

Fig. 4(c) shows the scaling behavior of $\rho_{xy}^{A}$ vs. $\rho_{xx}$. There is a jump at about 100 K corresponding to the temperature where the process of spin reorientation from the $c$ axis to the $ab$ plane is almost completed, as observed in the $\chi(T)$ curves. Using the formula $\rho_{xy}^{A}=\beta\rho_{xx}^{\alpha}$, the fits for $T\geq$ 100 K and $T<$ 100 K give $\alpha_{1}=$ 1.92(2) and $\alpha_{2}=$ 1.94(5), respectively. Both nearly quadratic relationships between $\rho_{xy}^{A}$ and $\rho_{xx}$ clearly indicates that the intrinsic KL or extrinsic side-jump mechanism dominates the AHE in Fe$_{3}$Sn$_{2}$ single crystal rather than extrinsic skew-scattering mechanism which will give the linear relationship between $\rho_{xy}^{A}$ and $\rho_{xx}$. By contrast, the Fe$_{3}$Sn$_{2}$ polycrystal exhibits the scaling behavior with much larger $\alpha$ (= 3.15),\cite{Kida} similar to those has also been observed in some inhomogeneous systems, such as Fe/Cr multilayers, Co-Ag granular films.\cite{Song, Xiong} Zhang\cite{ZhangSF} proposed that in the magnetic multilayered systems, the $\rho_{xy}^{A}$ will depend on the ratio of relaxation times or scattering potentials in magnetic and nonmagnetic layers in the long mean-free path limit where the mean-free path $\lambda$ is much larger than the layer thickness, and the scaling behavior with $\alpha=$ 2 will breakdown. This is one of possible explanations for the large $\alpha$ in Fe$_{3}$Sn$_{2}$ polycrystal when regarding the spin-disordered GBs and the gains of Fe$_{3}$Sn$_{2}$ as the nonmagnetic and magnetic layers, and the $\lambda$ is comparable or larger than the thickness of GB layer or gain size. By contrast, although the $\lambda$ of single crystal is larger than that of polycrystal, Fe$_{3}$Sn$_{2}$ single crystal can be regarded as a uniform magnet. It is more homogeneous than polycrystal without any heterostructure like GBs and the crystal size should be much larger than the $\lambda$. Thus, the local limit is valid and the anomalous Hall conductivity (AHC) $\sigma_{xy}^{A}$ ($\approx-\rho_{xy}^{A}/\rho_{xx}^{2}=-R_{s}\mu_{0}M/\rho_{xx}^{2}$) is independent of the scattering potential or the relaxation time, i.e., $\alpha=$ 2 for Fe$_{3}$Sn$_{2}$ single crystal.

Temperature dependence of $\sigma_{xy}^{A}$ is shown in Fig. 4(d). Above 100 K, the $\sigma_{xy}^{A}$ is about -170 $\Omega^{-1}$ cm$^{-1}$, insensitive to temperature. When $T$ is below 100 K, it increases quickly to about -400 $\Omega^{-1}$ cm$^{-1}$. This variation is also reflected in the scaling relationship between $\rho_{xy}^{A}$ and $\rho_{xx}$, leading to the jump of prefactor $\beta$ significantly because the $M$ is almost unchanged with temperature between 5 and 350 K. Theoretically, the intrinsic contribution of $|\sigma_{xy, in}^{A}|$ is of the order of $e^{2}/(h a)$, where $e$ is the electronic charge, $h$ is the Plank constant and $a$ is the lattice constant.\cite{Onoda} Taking $a=V^{1/3}\sim$ 12.8 \AA$\,$ approximately, the $|\sigma_{xy, in}^{A}|$ is $\sim$ 302 $\Omega^{-1}$ cm$^{-1}$, close to the value of $|\sigma_{xy}^{A}|$ at 5 K ($\sim$ 402 $\Omega^{-1}$ cm$^{-1}$). By contrast, the extrinsic side-jump contribution of $|\sigma_{xy, sj}^{A}|$ has been shown to be on the order of $(e^{2}/(h a)(\varepsilon_{SO}/E_{F})$, where $\varepsilon_{SO}$ and $E_{F}$ is the SOI and Fermi energy, respectively.\cite{Nozieres} Since the $\varepsilon_{SO}/E_{F}$ is usually less than 10$^{-2}$ for the metallic ferromagnets, the extrinsic side-jump contribution should be small and the AHE of Fe$_{3}$Sn$_{2}$ is dominated by the intrinsic Berry-phase KL contribution. It would be interesting to carry out the theoretical calculation to verify this analysis, as did for bcc Fe and SrRuO$_{3}$.\cite{Yao,FangZ} The dramatic increase of the $\sigma_{xy}^{A}$ at 100 K is unclear at present. Since the large contribution to $\sigma_{xy,in}^{A}$ comes from the enhanced Berry curvature near the avoided crossings of band dispersions at $E_{F}$ which only occur in very small regions of $k$ space, the $\sigma_{xy,in}^{A}$ is very sensitive to the topology of Fermi surface.\cite{Yao} The subtle changes of Fermi surface could happen due to the spin-reorientation transition below 100 K, leading to the variation of $\sigma_{xy}^{A}$. The decrease of $n_{a}$ also reflects this change of Fermi surface to some extent (inset of Fig. 4(a)).

For the intrinsic AHC, the $|\sigma_{xy, in}^{A}|$ is proportional to the $M$,\cite{ZengC,Manyala} the scaling coefficient $S_{H}=\mu_{0}R_{s}/\rho_{xx}^{2}=-\sigma_{xy}^{A}/M$ should be constant and independent of temperature. As shown in the inset of Fig. 4(d), when $T$ is away from 100 K, the $S_{H}$ is nearly constant, confirming the intrinsic KL mechanism is dominant contribution of AHE. The sudden change of $S_{H}$ at $\sim$ 100 K is related to the evolution of $|\sigma_{xy, in}^{A}|$. The value of $S_{H}$ is comparable with those in the traditional itinerant ferromagnets, such as Fe and Ni ($S_{H}\sim$ 0.01 - 0.2 V$^{-1}$).\cite{Dheer,Jan} Thus, even there is rather large $R_{s}$ in Fe$_{3}$Sn$_{2}$ single crystal in contrast to Fe and Ni, the larger $\rho_{xx}$ in the former leads to similar values of $S_{H}$ in these materials.

Previous theoretical studies predicted that an integer or fractional quantum Hall (IQH or FQH) states can appear at high temperature in the ferromagnet with the two-dimensional geometrically frustrated kagome lattice.\cite{Tang,Ohgushi} The IQH is corresponding to the fully filled flatband with a nonzero Chern number and the FQH state results from the 1/3 or 1/2 filling of that band. Tang et al. proposed that Fe$_{3}$Sn$_{2}$ is one of possible materials realizing these QH states.\cite{Tang} Present study indicates that instead of the appearance of QH states, the intrinsic AHE exists in Fe$_{3}$Sn$_{2}$. The metallic behavior and the different filling of the flatband in Fe$_{3}$Sn$_{2}$ from theoretically required 1/3 or 1/2 filling could be related to the absence of IQH or FQH states. Moreover, the finite interlayer interaction between kagome bilayer of Fe could also have some influences on the band gap $\Delta$, electron-electron interaction $U$ and band width $W$ and the condition of $\Delta\gg U\gg W$ favoring the FQH state may not be satisfied. Because the AHE can be regarded as the unquantized version of the QH effect once the Fermi energy is not lying in the gap,\cite{Nagaosa,XiaoD} both of the AHE observed in experiment and the QH states predicted by theory reflect the nontrivial topology of bands for the ferromagnetic system with geometrically frustrated kagome lattice.\cite{Ohgushi}

\section{Conclusion}

In summary, we successfully grow the Fe$_{3}$Sn$_{2}$ single crystal using the self-flux method. The Fe$_{3}$Sn$_{2}$ contains geometrically frustrated kagome bilayers of Fe and has a ferromagnetic transition at about 640 K followed by another spin-reorientation transition from $c$ axis to $ab$ plane at the lower temperature. The transport and magnetic measurement indicates that Fe$_{3}$Sn$_{2}$ single crystal exhibits a scaling relationship between $\rho_{xy}^{A}$ and $\rho_{xx}$, $\rho_{xy}^{A}\propto\rho_{xx}^{2}$. The detailed analysis suggests that the AHE should be dominated by the intrinsic KL mechanism and there is a large $\sigma_{xy}^{A}$. Further theoretical calculation is highly expected in order to clarify whether such large $\sigma_{xy}^{A}$ is related to the topologically nontrivial bands in Fe$_{3}$Sn$_{2}$, which originates from the ferromagnetism with geometrically frustrated kagome bilayer of Fe. Moreover, although the QH states are absent in the undoped Fe$_{3}$Sn$_{2}$, adjusting the band filling by doping could be a way to realize the theoretically predicted QH states. On the other hand, the $\sigma_{xy}^{A}$ exhibits a sudden change at about 100 K, corresponding to the ending temperature of spin-reorientation transition. This jump of $\sigma_{xy}^{A}$ is tentatively explained by the evolution of Fermi surface.

\section{Acknowledgments}

This work was supported by the Ministry of Science and Technology of China (2012CB921701, 2016YFA0300504), the Fundamental Research Funds for the Central Universities, and the Research Funds of Renmin University of China (RUC) (15XNLF06, 15XNLQ07), and the National Natural Science Foundation of China (Grant No. 11574394).


\begin{thebibliography}{99}

\bibitem{Hall} E. H. Hall, Philos. Mag. \textbf{10}, 301 (1880); \textbf{12}, 157(1881).

\bibitem{Nagaosa} N. Nagaosa, J. Sinova, S. Onoda, A. H. MacDonald, N. P. Ong, Rev. Mod. Phys. \textbf{82} 1539 (2010).

\bibitem{Karplus} R. Karplus and J. M. Luttinger, Phys. Rev. \textbf{95}, 1154 (1954).

\bibitem{Jungwirth} T. Jungwirth, Q. Niu, and A. H. MacDonald, Phys. Rev. Lett. \textbf{88}, 207208 (2002);

\bibitem{Onoda1} M. Onoda and N. Nagaosa, Phys. Rev. Lett. \textbf{90}, 206601 (2003).

\bibitem{Smit} J. Smit, Physica \textbf{21}, 877 (1955); \textbf{24}, 39 (1958).

\bibitem{Berger} L. Berger, Phys. Rev. B \textbf{2}, 4559 (1970).

\bibitem{Taguchi} Y. Taguchi, Y. Oohara, H. Yoshizawa, N. Nagaosa, and Y. Tokura, Science \textbf{291} 2573 (2001).

\bibitem{Machida} Y. Machida, S. Nakatsuji, S. Onoda, T. Tayama, and T. Sakakibara, Nature \textbf{463}, 210 (2010).

\bibitem{Nakatsuji} S. Nakatsuji, N. Kiyohara, and T. Higo, Nature \textbf{527}, 212 (2015).

\bibitem{Malaman} B. Malaman, D. Fruchart, and G. L. Ca\"{e}r, J. Phys. F: Met. Phys. \textbf{8} 2389 (1978).

\bibitem{Caer1} G. L. Ca\"{e}r, B. Malaman, and B. Roques, J. Phys. F: Met. Phys. \textbf{8} 323 (1978).

\bibitem{Caer2} G. L. Ca\"{e}r, B. Malaman, L. H\"{a}ggstr\"{o}m, and T. Ericsson, J. Phys. F: Met. Phys. \textbf{9} 1905 (1979).

\bibitem{Fenner} L. A. Fenner, A. A. Dee, and A. S. Wills, J. Phys.: Condens. Matter \textbf{21}, 452202 (2009).

\bibitem{Kida} T. Kida, L. A. Fenner, A. A. Dee, I. Terasaki, M. Hagiwara, and A. SWills, J. Phys.: Condens. Matter \textbf{23}, 112205 (2011).

\bibitem{Tang} E. Tang, J. W. Mei, and X. G. Wen, Phys. Rev. Lett. \textbf{106}, 236802 (2011).

\bibitem{Volkenshtein} N. V. Volkenshtein and G. V. Fedorov, Sov. Phys.-JETP \textbf{11}, 48 (1960).

\bibitem{Kaul} S. N. Kaul, Phys.Rev. B \textbf{20}, 5122 (1979).

\bibitem{Hurd} C. M. Hurd, in The Hall effect and its Applications, edited by C. Chien and C. R. Westgate (Plenum, New York, 1980).

\bibitem{Song} S. N. Song, C. Sellers, and J. B. Ketterson, Appl. Phys. Lett. \textbf{59} 479 (1991).

\bibitem{Xiong} P. Xiong, G. Xiao, J. Q. Wang, J. Q. Xiao, J. S. Jiang, and C. L. Chien, Phys. Rev. Lett. \textbf{69}, 3220-3223 (1992).

\bibitem{ZhangSF} S. Zhang, Phys. Rev. B \textbf{51}, 3632 (1995).

\bibitem{Onoda} S. Onoda, N. Sugimoto, and N. Nagaosa, Phys. Rev. Lett. \textbf{97}, 126602 (2006).

\bibitem{Nozieres} P. Nozi\`{e}res and C. Lewiner, J. Phys. (Paris) \textbf{34}, 901 (1973).

\bibitem{Yao} Y. Yao, L. Kleinman, A. H. MacDonald, J. Sinova, T. Jungwirth, D.-S. Wang, E. Wang, and Q. Niu, Phys. Rev. Lett. \textbf{92}, 037204 (2004).

\bibitem{FangZ} Z. Fang, N. Nagaosa, K. S. Takahashi, A. Asamitsu, R. Mathieu, T. Ogasawara, H. Yamada, M. Kawasaki, Y. Tokura, and K. Terakura, Science \textbf{302}, 92 (2003).

\bibitem{ZengC} C.  Zeng,  Y. Yao,  Q.  Niu,  and  H.  H. Weitering, Phys.  Rev.  Lett. \textbf{96}, 037204 (2004).

\bibitem{Manyala} N. Manyala, Y. Sidis, J. F. Ditusa, G. Aeppli, D. P. Young, and Z. Fisk, Nat. Mater. \textbf{3}, 255 (2004).

\bibitem{Dheer} P. N. Dheer, Phys. Rev. \textbf{156}, 637 (1967).

\bibitem{Jan} J.-P. Jan and H. M. Gijsman, Physica \textbf{18}, 339 (1952).

\bibitem{Ohgushi} K. Ohgushi, S. Murakami, and N. Nagaosa, Phys. Rev. B \textbf{62}, R6065 (2000).

\bibitem{XiaoD} D. Xiao, M.-C. Chang, Q. Niu, Rev. Mod. Phys. \textbf{82} 1959 (2010).

\end{thebibliography}
\end{document}